\documentclass[12pt]{article}

\usepackage{makeidx}
\usepackage{epsfig}
\usepackage{cite}
\usepackage{color}
\usepackage{graphicx}
\usepackage{xspace}
\makeindex

\usepackage{amsmath}
\usepackage{amsthm}
\usepackage{amscd}
\usepackage{amsfonts}
\usepackage[psamsfonts]{amssymb}

\def\Matter{\bf m}

\begin{document}
\begin{titlepage}

\begin{flushright}
UCB-PTH-07/20\\
LBNL-63518
\end{flushright}

\vfil\

\begin{center}
{\Large{\bf Non-Relativistic Superstring Theories}}
\vfil

Bom Soo Kim

\bigskip

Department of Physics, University of California, Berkeley, CA 94720\\
Theoretical Physics Group, Lawrence Berkeley National Laboratory, Berkeley, CA 94720\\
{\it bskim@socrates.berkeley.edu}
\vfil

\end{center}

\begin{abstract}

We construct a supersymmetric version of the ``critical'' non-relativistic bosonic string theory\cite{Kim:2007hb} 
with its manifest global symmetry. 
We introduce the anticommuting $bc$ CFT which is the super partner of the $\beta\gamma$ CFT. The conformal 
weights of the $b$ and $c$ fields are both $1/2$. The action of the fermionic sector can be transformed into 
that of the relativistic superstring theory. We explicitly quantize the theory with manifest $SO(8)$ symmetry 
and find that the spectrum is similar to that of Type IIB superstring theory. There is one notable 
difference: the fermions are non-chiral. We further consider ``noncritical'' generalizations of the 
supersymmetric theory using the superspace formulation. There is an infinite range of possible string theories similar to 
the supercritical string theories. We comment on the connection between the critical non-relativistic string theory and 
the lightlike Linear Dilaton theory.

\end{abstract}

\vspace{0.5in}

\end{titlepage}

\section{Introduction}

Time-dependent backgrounds in string theory are hard to analyze\cite{Liu:2002ft}. Perturbative string theory 
breaks down in some spacetime regions due to a large string coupling, and it appears that a full 
nonperturbative string theory formulation is required.
One clean example with the lightlike Linear Dilaton theory is proposed in \cite{Craps:2005wd}.
On the other hand, there are some interesting developments which emphasize the role of perturbative string theory 
in the analysis of time-dependent backgrounds\cite{McGreevy:2005ci, Horava:2007yh}. 
But the complete understanding of time-dependent backgrounds is still out of reach in string theory. 

It turns out that many interesting cosmological solutions have broken Lorentz symmetry. And it is interesting to 
consider these solutions with their manifest global symmetry. Furthermore fundamental issues related to time, 
especially to ``emergent time'', is not clear (see, {\it e.g.,}\cite{Seiberg:2006wf}). 
Thus it is interesting to consider alternative approaches, which can shed light on time-dependent backgrounds 
and on fundamental issues of time.  

Recently a bosonic string theory with manifest Galilean symmetry in target space was constructed in an elementary 
fashion\cite{Kim:2007hb}, motivated by earlier works\cite{Gomis:2000bd, Danielsson:2000gi, Danielsson:2000mu}. 
These non-relativistic string theories clearly treat time differently than relativistic string theory.
In non-relativistic string theories, time in target space can be described 
by the first order nonunitary $\beta\gamma$ CFT, while second order $X^0$ CFT plays the role of time in the relativistic theory. 
Thus we can hope to obtain some insights on the issues of time-dependent backgrounds in string theory 
from this very different approach. As we mention in the final section of this paper, there are some intriguing 
pieces of evidence that 
these non-relativistic string theories can be connected to known time-dependent backgrounds in string theory. 
This possibility opens up a new framework for addressing the issues related to time and to time-dependent string solutions.

With these motivations, we briefly review the construction 
of the bosonic non-relativistic string theory, which has a manifest Galilean symmetry in target space. 
Compared to earlier works, 
the theory does not assume a compact coordinate and has a simpler action, a $\beta\gamma$ CFT in addition to the 
usual bosonic $X$ CFTs. The first order $\beta\gamma$ CFT is directly related to time and energy in target space. 
Time in target space is parametrized by a one parameter family of selection sector and is explicitly realized through 
the generalized Galilean boost symmetry of the action. We quantize the theory in an elementary fashion which 
reveals many interesting features. The spectrum is very similar to the relativistic bosonic string theory, except for the 
overall motion of the string which is governed by a non-relativistic energy dispersion relation.
The ground state has the energy 
\begin{equation}
E = \frac{1}{p p'}\left(\frac{\alpha'}{4} k^ik_i - 1\right)\ ,
\end{equation}
where $p$ and $p'$ are the parameters which specify the selection sector and the ground state vertex operator, 
respectively, and $k^i$s are the transverse momenta. 
The particle corresponding to the ground state is still ``tachyonic'' because it is possible to have negative energy 
for the range $\frac{\alpha'}{4} k^ik_i \leq 1$. Thus it is desirable to remove this state from the spectrum. 
The first excited state has $24$ degrees of freedom which transform into each other under $SO(24)$ rotations. 

The world sheet constraint algebra imposes strong restrictions on the spectrum of string theories.
We can enlarge the world sheet constraint algebra by adding the supercurrents to construct non-relativistic 
superstring theories. We start with the non-relativistic superstring action in terms of the component fields in the 
critical case, which reveals an interesting simplification in the fermionic sector. The fermionic sector can be 
rewritten in the same form as in the relativistic superstring theory with a simple transformation. 
The rest of the quantization is very similar 
to that of the relativistic superstring theory, except for a different global symmetry structure. We explicitly 
construct the vertex operators using the bosonization technique,then we quantize the theory and check the modular 
invariance. We encounter a non-relativistic analogue of the 
Dirac equation in the ground state of the $R$ sector. By solving the equation we show that the fermionic 
sector has eight physical degrees of freedom which transform in the spinor representation ${\bf 8}$ of $SO(8)$. 
But there is one clear difference: the fermions in this theory are non-chiral. We contrast this to the 
relativistic case. This is done in section 2. 

In section 3, we consider the ``noncritical'' version of non-relativistic superstring theories. 
We present the superspace formulation of the new first order matter ${\bf \Sigma\Gamma}$ CFT in detail. 
There exist an infinite range of possible string theories for the general conformal 
weights of the ${\bf \Sigma\Gamma}$ CFT. There are two different categories in the noncritical theories distinguished 
by the conformal weight of the $\beta\gamma$ CFT: those with integer weight and those with half 
integer conformal weight. The former case is similar to the case we quantize in this paper. 
The latter case seems more exotic and it is expected to give us a rather different view on the 
geometric interpretation of target space. 

Using the world sheet constraint algebra, we construct all possible string theories with 
extended supersymmetry in section 4. The bosonic and supersymmetric non-relativistic string cases 
are presented here. We comment on some immediate observations. We conclude in section 5. 
In section 6, we mention possible intriguing 
applications of this non-relativistic string theory to time-dependent string backgrounds such as 
the lightlike Linear Dilaton theory.

\section{Critical Non-Relativistic Supersymmetric String}

\subsection{New Matter $\beta\gamma$ CFT and $bc$ CFT}

We start with a full non-relativistic superstring action of component fields in the conformal gauge 
\begin{eqnarray}
S &=&  \int \frac{d^2z}{2\pi} \left( \beta \bar{\partial} \gamma + \bar{\beta} \partial \bar{\gamma} 
+  \frac{1}{\alpha'} \partial X^i \bar{\partial} X_i + b_g \bar{\partial} c_g 
+ \bar{b_g} \partial \bar{c_g}\right) \nonumber  \\ 
&+& \int \frac{d^2z}{2\pi} \left( b \bar{\partial} c + \bar{b} \partial \bar{c} 
+  \frac{1}{2} \left( \psi^i \bar{\partial} \psi_i + \bar{\psi}^i \partial \bar{\psi}_i \right) 
+ \beta_g \bar{\partial} \gamma_g + \bar{\beta_g} \partial \bar{\gamma_g}\right) \ ,
\label{fermionicaction1}
\end{eqnarray}
where $i$ runs from 2 to 9 for $X^i$ and $\psi^i$ for the critical non-relativistic superstring theory. 
The commuting matter $\beta\gamma$ CFT has weights, $h(\beta) = 1$ and $h(\gamma) = 0$, and has 
its central charge, $c_{\beta\gamma} = 2$. 
The anticommuting matter $bc$ CFT, whose central charge is $c_{bc} = 1$, 
has weight $h(b) = 1/2$ and $h(c) = 1/2$. 
In conventional notation for the superstring case, the total central charge of the matter sector is 
$\hat{c}^{\bf m} = \frac{2}{3} c^{\bf m} = \frac{2}{3} (3 + \frac{3}{2} D)$, which cancels the central charge from the 
ghost sector $\hat{c}^{\bf gh} = \frac{2}{3} c^{\bf gh} = \frac{2}{3} (-26 + 11) = -10$. 
Thus this theory is anomaly free if $D = 8$. This is indicated above by the spatial index $i$ which runs from 2 to 9. 
We consider the cases with general conformal weights in the matter $\beta\gamma$ 
and $bc$ CFTs in the next section. The case with conformal weight of $\beta$ as $1$ 
is rather special and we will call it as the ``critical'' case as in bosonic non-relativistic theory.

We briefly comment on the new matter $\beta\gamma$ and $bc$ CFTs. Their OPEs are 
\begin{eqnarray}
\gamma(z_1) \beta(z_2)\hspace{0.1 in} \sim & \frac{1}{z_{12}} &\sim \hspace{0.1 in} - \beta(z_1) \gamma(z_2) \ , 
\\ 
b(z_1) c(z_2) \hspace{0.1 in} \sim & \frac{1}{z_{12}} &\sim \hspace{0.1 in} c(z_1) b(z_2) \ .
\end{eqnarray} 
The bosonic and the fermionic energy momentum tensors and their mode expansions are 
\begin{eqnarray}
T_{b}^{\beta\gamma b c} =  - (\partial \gamma) \beta -\frac{1}{2} c (\partial b) + \frac{1}{2}(\partial c) b 
&=& \sum_{m \in {\bf Z}} \frac{L_m}{z^{m+2}} \ , \label{bosonicenergymomentumtensor111}\\
T_{f}^{\beta\gamma b c} =  \frac{1}{2} c \beta - \frac{1}{2} (\partial \gamma) b 
&=& \sum_{r \in {\bf Z} + \nu} \frac{G_r}{2 \cdot z^{r+3/2}} \ . \label{bosonicenergymomentumtensor222} 
\end{eqnarray}
As is well known there are two possible sectors for the fields with the half integer conformal weight.  
These are $\nu = 0$ and $\nu = 1/2$ cases corresponding to $R$ sector and $NS$ sector, respectively. 
We can also find mode expansions and their hermiticity properties of the fields
\begin{eqnarray}
\gamma(z) = \sum_{n \in {\bf Z}} \frac{\gamma_n}{z^{n}} \ , \hspace{0.3 in} \gamma_n^{\dagger} =  \gamma_{-n} \ , 
&& \beta(z) = \sum_{n \in {\bf Z}} \frac{\beta_n}{z^{n+ 1}} \ , \hspace{0.3 in} \beta_{n}^{\dagger} =  - \beta_{-n} \ , \\
c(z) = \sum_{r \in {\bf Z} + \nu}  \frac{c_r}{z^{r+1/2}} \ , \hspace{0.2 in} c_r^{\dagger} =  c_{-r} \ ,
&& b(z) = \sum_{r \in {\bf Z} + \nu}  \frac{b_r}{z^{r+ 1/2}}  \ , \hspace{0.2 in} b_{r}^{\dagger} =  b_{-r} \ .
\end{eqnarray}
And the mode expansions for the energy momentum tensors are 
\begin{eqnarray}
&& L_m^{\beta\gamma b c} = \sum_{n \in {\bf Z}} n \beta_{m-n}\gamma_n + \sum_{s \in {\bf Z} +\nu} (s - m/2) b_{m-s} c_s 
+ a\delta_{m,0} \ , \\
&& G_r^{\beta\gamma b c} = \sum_{m \in {\bf Z}} \left( c_{r-m} \beta_m + m \gamma_m b_{r-m} \right) \ .
\end{eqnarray}
There is a normal ordering constant for $L_0$ in each sector 
\begin{equation}
a_R^{\beta\gamma b c} = \frac{1}{8} \ , \hspace{0.3 in} a_{NS}^{\beta\gamma b c} = 0 \ .
\end{equation} 
This is only from the new matter sector, $\beta\gamma$ and $ b c$ CFTs, and is one part of the total normal 
ordering constant.\footnote
{It is important to observe that the total normal ordering constant for non-relativistic 
superstring theory is the same as that of the relativistic theory 
\begin{equation}
a_R = 0, \hspace{0.3 in} a_{NS} = -\frac{1}{2} \ , \nonumber
\end{equation} 
because there are other contributions from the $X^i$ CFTs and the ghost CFT.
}

\subsection{Fermionic Sector and its Symmetry}

The fermionic $bc$ CFT is a new ingredient of this non-relativistic superstring theory. 
There are immediate observations which are rather interesting. 
As we briefly mentioned at the beginning of this section, the conformal weights of the fields $b$, $c$ and 
all the other fermionic fields $\psi^i$ are equal and the value is $1/2$. From this observation, 
we can think about a transformation 
\begin{equation}
c = \frac{1}{\sqrt{2}}(\psi^1 - \psi^0) \ , \qquad 
b = \frac{1}{\sqrt{2}}(\psi^1 + \psi^0) \ .
\end{equation} 
Combining these fields with the other fermionic fields $\psi^i$, we can see that the action of the 
fermionic sector is exactly the same as that of the relativistic one 
\begin{equation}
S_F = \int \frac{d^2z}{2\pi} \left( b \bar{\partial} c + \bar{b} \partial \bar{c} 
+  \frac{1}{2} \left( \psi^i \bar{\partial} \psi_i + \bar{\psi}^i \partial \bar{\psi}_i \right) \right)
= \int \frac{d^2z}{4\pi}
\left( \psi^\mu \bar{\partial} \psi_\mu + \bar{\psi}^\mu \partial \bar{\psi}_\mu \right) \ ,
\end{equation}
where $\mu$ runs from 0 to 9. 
We can naively think that there are $SO(9,1)$ invariance in the fermionic sector of this non-relativistic 
superstring theory. But as is obvious from the original action, there is no symmetry transformation 
which connects the fields $\psi^0, \psi^1$ 
and the other transverse fields $\psi^i$. The symmetry groups of the fermionic sector 
are the $SO(8)$ rotations among the $\psi^i$s as well as a one-parameter family of superconformal symmetry which is related 
to rescaling $\beta \rightarrow x \beta$ and $\gamma \rightarrow \gamma /x$.\footnote{We realize that there 
exist this symmetry when we have discussions with Professor Ori Ganor and with Professor Ashvin Vishwanath. 
We thank for their questions and comments related to this.}
 The latter is actually 
realized as the relative rescaling between $k^\gamma$ and $p'$ in the bosonic string case, related  
by rescaling $k^\gamma \rightarrow x k^\gamma$ and $p' \rightarrow p' /x$. We can denote this zero dimensional 
conformal symmetry as ``$SO(1,1)$'', thus the symmetry group turns out to be $SO(1,1) \times SO(8)$. 
This symmetry group becomes important when we consider a non-relativistic analogue of the Dirac equation. 
Even though we know there is no relativistic $SO(9,1)$ symmetry, we still use the relativistic notation 
to make the expression simple and to get some intuitions from the relativistic results.

\subsection{Vertex Operators}

Most of the vertex operators for this theory are already known. The vertex operators of 
the $X^i,\psi^i$ CFTs and of the superconformal ghost sector with the $b_g c_g$ and $\beta_g \gamma_g$ CFTs 
are already well understood and can be found in many places (see, {\it e.g.,} \cite{fms, pol, gsw}).
Constructing vertex operators for the bosonic $\beta\gamma$ CFT is considered in \cite{Kim:2007hb, Gomis:2000bd}. 

Thus let's concentrate on the vertex operators of the fermionic $bc$ CFT. 
The fermionic matter sector, in terms of the fermionic fields $\psi^\mu$, $\mu = 0 \cdots 9$, 
has well understood vertex operators in the relativistic string theory\cite{fms, gsw, pol}. 
Thus we can just borrow the results from them with caution. In this section we will use both the notations 
$\psi^0, \psi^1$ and $bc$.

For the Neveu-Schwarz ($NS$) sector, there is no $r=0$ mode and we can define the ground state 
to be annihilated by all $r>0$ modes 
\begin{equation}
\psi_r^{\mu} ~ | 0; k^\gamma, k^{\bar{\gamma}}, \vec{k} \rangle_{NS} = 0 \ , \hspace{0.3 in}  r > 0 \ . 
\end{equation} 
This ground state is ``tachyonic''. The vertex operator corresponding to $NS$ ground state is 
\begin{eqnarray}
&&V_{NS, 0}(k^\gamma, k^{\bar{\gamma}}, k^i; z, \bar{z}) 
= e^{-\varphi} V_{0}(k^\gamma, k^{\bar{\gamma}}, k^i; z, \bar{z}) \ , \\
&&V_{0}(k^\gamma, k^{\bar{\gamma}}, k^i; z, \bar{z}) = g: e^{ik^\gamma \gamma + ik^{\bar{\gamma}} 
\bar{\gamma} - ip' \int^{z} \beta - iq' \int^{\bar{z}} \bar{\beta} + ik^i \cdot X_i} : \ ,  \label{groundvertex}
\end{eqnarray}
where the field $\varphi$ comes from the bosonization of the superconformal ghost fields and has nothing to do with 
the selection parameter $\phi$. And the bosonic ground state vertex operator $V_0$ was considered in 
\cite{Gomis:2000bd, Kim:2007hb} with  $k^\gamma$, $k^{\bar{\gamma}}$ and $k^i$ representing the overall continuous momenta 
along the coordinates $\gamma$, $\bar{\gamma}$ and $X^i$, respectively.

The first excited state in the $NS$ sector is a linear combination of the fermionic excitations 
$b_{-1/2}$, $c_{-1/2}$ and $\psi^i_{-1/2}$. 
\begin{equation}
| e ; k^\gamma, k^{\bar{\gamma}}, \vec{k}\rangle_{NS} 
= \left( e_c c_{-1/2} + e_b b_{-1/2} + e_i \psi_{-1/2}^{i} \right) 
| 0; k^\gamma, k^{\bar{\gamma}}, \vec{k} \rangle_{NS} \ . 
\end{equation} 
We use two different notations for the fermionic sector (i) $e_\mu\psi^\mu$ with $\mu = 0, \cdots , 9$ and 
(ii) $e_M\psi^M_{-1/2} = \left( e_c c_{-1/2} + e_b b_{-1/2} + e_i \psi_{-1/2}^{i} \right) $ with $i = 2, \cdots, 9$.
The vertex operator corresponding to the first excited state 
$V_{NS, 1}(k^\gamma, k^{\bar{\gamma}}, k^i; z, \bar{z})$ is 
\begin{eqnarray}
 e^{-\varphi} \psi^M V_{0}(k^\gamma, k^{\bar{\gamma}}, k^i; z, \bar{z})  & \hbox{or} &
e^{-\varphi} \psi^\mu V_{0}(k^\gamma, k^{\bar{\gamma}}, k^i; z, \bar{z}) \ . 
\end{eqnarray}
The modes with $r<0$ for the fields $\psi_r$ act as raising operators and each mode can be excited only once. 

The Ramond ($R$) sector ground state is degenerate due to the zero modes $\psi_0^{\mu}$ (or $\psi_0^{M}$). 
We can define the $R$ ground state to be those that are annihilated by all $r>0$ modes. 
And the zero modes satisfy the Dirac gamma matrix algebra with $\Gamma^{\mu} \cong \sqrt{2} \psi_0^{\mu}$. 
Since $\{\psi_r^{\mu}, \psi_0^{\nu} \} = 0$ for $r>0$, the zero modes $\psi_0^{\mu}$ take 
ground states into ground states. Thus the ground states form a representation of the gamma 
matrix algebra. For critical case with ``10 dimensions'' we can represent this as 
$| {\bf s} \rangle =| s_0\rangle \times | \vec{s}\rangle= | s_0\rangle \times | s_1, s_2, s_3, s_4 \rangle$ 
with $s_0, s_a = \pm 1/2$. 
Here we separate $s_0$ from the others to indicate that there is no symmetry transformation between $s_0$ and $\vec{s}$.  

It is convenient to combine two fermions, $\psi^2$ and $\psi^3$ for example, into a complex pair, 
$\psi \equiv \frac{1}{\sqrt{2}}(\psi^2 + i \psi^3)$ and $\psi^{\dagger} \equiv \frac{1}{\sqrt{2}}(\psi^2 - i \psi^3)$,\footnote{
Note that we use different notation for the complex field compared to \cite{pol}.} 
to consider a more general periodicity condition 
\begin{equation}
\psi (w + 2\pi) = e^{2\pi i \nu} \psi(w) \ ,
\end{equation}
for any real $\nu$. Here we concentrate on two cases $\nu = 0$ and $\nu = 1/2$. The mode expansions are  
\begin{equation}
\psi(z) = \sum_{r \in {\bf Z} +\nu} \frac{\psi_r}{z^{r+1/2}}, \qquad 
\psi^{\dagger}(z) = \sum_{s \in {\bf Z} -\nu} \frac{\psi^{\dagger}_s}{z^{s+1/2}} \ , \label{complexfermions11}
\end{equation}
with a commutation relation $\{ \psi_r, \psi^{\dagger}_s \} = \delta_{r, -s}$. 

We can define a reference state 
$|0 \rangle_{\nu}$ by 
\begin{equation}
\psi_{n+\nu} |0 \rangle_{\nu} = \psi^{\dagger}_{n + 1 -\nu} |0 \rangle_{\nu} = 0 \ , \qquad n = 0, 1, \cdots \ .
\end{equation}
The first nonzero terms in the Laurent expansions are related to the indices $r = -1 +\nu$ and $s = - \nu$. 
And these conditions can uniquely identify the state $|0 \rangle_{\nu}$. Similarly for the corresponding vertex 
operator ${\cal A}_\nu$, the OPEs
\begin{equation}
\psi(z) {\cal A}_\nu (0) = {\cal O}(z^{-\nu + 1/2}), \qquad 
\psi^{\dagger}(z) {\cal A}_\nu (0) = {\cal O}(z^{\nu - 1/2}) \ 
\end{equation}
can determine the vertex operator as 
\begin{equation}
{\cal A}_\nu \simeq e^{i(-\nu + 1/2 )H} \ . \label{bosonizedvertex}
\end{equation}
This vertex operator has weight $h = \frac{1}{2} (\nu-\frac{1}{2})^2$. The boundary conditions are same 
for $\nu$ and $\nu + 1$, but the reference states are not. The reference state is a ground state only for 
$ 0 \leq \nu \leq 1$. For the $R$ sector with $\nu = 0$, there are two degenerate ground states which can be identified as 
$|s \rangle \cong e^{is H}$ with $s =  1/2$ and $s = -1/2$. 

It is convenient to use bosonization to take care of branch cut which arises in the fields with the 
half integer conformal weight. And the explicit bosonization expressions are 
\begin{eqnarray}
\frac{1}{\sqrt{2}}(\psi^1 - \psi^0) = c \cong e^{-iH^0} \ ,
&&\frac{1}{\sqrt{2}}(\psi^1 + \psi^0) = b \cong e^{iH^0} \ , \\
\frac{1}{\sqrt{2}}(\psi^{2a} \pm i \psi^{2a+1} ) \cong e^{\pm i H^a} \ ,  &&a = 1, \cdots, 4 \ , 
\end{eqnarray}
where $H(z)$ fields are the holomorphic part of corresponding scalar fields. 
Then the corresponding vertex operator $\Theta_{\bf s}$ for an $R$ sector ground state 
$|{\bf s} \rangle = | s_0, \vec{s} \rangle $ is
\begin{equation}
\Theta_{\bf s} \cong \exp \Big[ i s_0 H^0\Big]\times \exp \Big[ i\sum_{a=1}^{4} s_a H^a\Big].
\end{equation}
This spin field operator produces a branch cut in $\psi^{\mu}$ and need to be combined with 
an appropriate antiholomorphic vertex operator. 

Thus the $R$ ground state vertex operators are
\begin{equation}
V_{R, 0}(s_0, \vec{s}; k^\gamma, k^{\bar{\gamma}}, k^i; z, \bar{z}) = 
e^{-\varphi/2} \Theta_{\bf s} V_{0}(k^\gamma, k^{\bar{\gamma}}, k^i; z, \bar{z}) \ ,
\end{equation}
where $\varphi$ is related to the bosonization of the superconformal ghost fields and $V_0$ is given in equation 
(\ref{groundvertex}). Now we are ready to quantize the theory.

\subsection{Quantization}

In the old covariant quantization procedure, we ignore the ghost excitations and concentrate 
on the matter sector, which has the $X^i$, $\psi^i$, $\beta\gamma$ and $ b c$ CFTs. 
We impose the physical states conditions 
\begin{eqnarray}
\left(L_{n}^{\Matter} + a \delta_{n,0} \right) |\psi\rangle = 0, \hspace{0.1 in} n \geq 0 \ , &&
\hspace{0.2 in} G_r^{\Matter} | \psi \rangle = 0, \hspace{0.1 in} r \geq 0 \ ,
\end{eqnarray}
where '$\Matter$' denotes the matter sector. We can construct spurious states which are orthogonal to all 
physical states such as 
\begin{eqnarray}
L_{n}^{\Matter} | \chi\rangle \ , \hspace{0.2 in} n < 0 \ , &\hspace{0.2 in}&
\hspace{0.2 in} G_r^{\Matter} | \chi \rangle \ , \hspace{0.2 in} r < 0 \ .
\end{eqnarray}
These states satisfy $\langle \psi | L_n^{\bf m} |\chi \rangle = 0$ and 
$ \langle \psi |G_r^{\Matter} | \chi \rangle = 0 $. If these states satisfy the physical 
state conditions, then we call them null states. We need to impose equivalence relations to get a physical 
Hilbert space.

\bigskip
{\it $NS$ sector}

The $NS$ sector with $\nu = 1/2$ is simpler and we consider this first. 
For the ground state (with simplified notation $|0; k\rangle_{NS}$ instead of 
$|0; k^\gamma, k^{\bar{\gamma}}, \vec{k} \rangle_{NS}$), 
the physical state condition $\left( L_0^{\Matter} -\frac{1}{2} \right) |0; k\rangle_{NS}  = 0$ 
gives us the mass shell equation  
\begin{equation}
\frac{\alpha'}{4} \vec{k}^2 - k^\gamma p' -\frac{1}{2} = 0 \ .
\end{equation}   
The other physical state conditions, $L_n^{\Matter} |0; k\rangle_{NS} = 0$ for $ n > 0$ and 
$G_{r}^{\Matter} |0; k\rangle_{NS} = 0$ for $ r \geq 1/2$, are trivial. Thus there is one equivalence class, 
corresponding to a scalar particle.

The first excited level (with simplified notation $|e; k \rangle_{NS}$ instead of 
$|e; k^\gamma, k^{\bar{\gamma}}, \vec{k} \rangle_{NS}$) has 10 states 
\begin{equation}
|e; k \rangle_{NS} = \left(e_c c_{-1/2}  + e_b b_{-1/2}  +  e_i \psi_{-1/2}^i \right) |0; k \rangle_{NS}.
\end{equation}
The nontrivial physical state conditions, $\left( L_0^{\Matter} -\frac{1}{2} \right) |e; k \rangle_{NS}=0$ 
and $G_{1/2}^{\Matter} |e; k \rangle_{NS}=0$, give us 
\begin{eqnarray}
\frac{\alpha'}{4} \vec{k}^2 - k^\gamma p' = 0 \ , \label{massshellforns1} \\ 
- p' e_c + k^\gamma e_b + (\alpha'/2)^{1/2} k^i e_i = 0 \ ,
\end{eqnarray}   
while a spurious state 
\begin{equation}
G_{-1/2}^{\Matter} |0; k \rangle_{NS} 
= \left( (\alpha'/2)^{1/2} k^i \psi_{i,1/2} +  k^\gamma c_{-1/2} - p' b_{-1/2}\right) |0; k \rangle_{NS}
\end{equation}
is physical and null. Thus there is an equivalent relation 
\begin{equation}
\left(e_c, \hspace{0.1 in} e_b, \hspace{0.1 in}e_i \right) \cong 
\left(e_c + k^\gamma,\hspace{0.1 in} e_b -p' ,\hspace{0.1 in} e_i+ (\alpha'/2)^{1/2} k_i \right) \ .
\end{equation}
Thus for the first excited state in the $NS$ sector, there are only 8 independent degrees of freedom. 

The global symmetries are the conformal scaling and the $SO(8)$ rotation, $SO(1,1) \times SO(8)$, 
as we point out above. 
At this stage, these symmetries are manifest in the equation (\ref{massshellforns1}).
But we show in the previous work\cite{Kim:2007hb} that the energy dispersion relation for the particle 
corresponding to this level is actually 
\begin{equation}
E = p_t = \frac{1}{pp'} \left(\frac{\alpha'}{4} \vec{k}^2 - 1 \right) \ ,
\end{equation}
where $p$ and $p'$ are parameters specifying a selection sector and the ground state vertex operator, respectively.
Thus non-relativistic particles have $SO(8)$ symmetry which is smaller than $SO(1,1) \times SO(8)$. 
The explicit dependence of energy on the parameter $p'$ breaks $SO(1,1)$ scaling symmetry. Particularly, 
at the first excited level of $NS$ sector, these eight degrees of freedom transform into each other 
in the vector representation of ${\bf 8}_v$ of $SO(8)$ similar to the case of relativistic massless 
excitations.\footnote{\label{so11spectrumfootnote}
There is another way to think about the expression (\ref{massshellforns1}). 
Rather than breaking $SO(1,1)$ symmetry, we can go to a frame, $k^i = 0 $ for $i = 2, \cdots , 8$ and $k^9 \neq 0$, 
which is similar to the relativistic consideration and keeps the $SO(1,1) \times SO(7)$ symmetry. 
For further explanation, please see the appendix. }

\bigskip 
{\it $R$ sector}

In the $R$ sector, we have degenerate ground states $|v, u; k \rangle_R = 
|s_0, \vec{s}; k\rangle_R ~ \left( v_{s_0} \otimes u_{\vec{s}} \right)$, where $v$ and $u$ are 
``polarizations'' along $bc$ and $\psi^i$, respectively. The nontrivial physical conditions are 
\begin{eqnarray}
&&0 = L_0^{\Matter} |v, u; k \rangle_R = \left(\frac{\alpha'}{4} \vec{k}^2 - k^\gamma p' \right)
|v, u; k \rangle_R \ , \\
&&0 = G_0^{\Matter} |v, u; k \rangle_R = \left(\left(\frac{\alpha'}{2}\right)^{1/2} k^i \psi_{0, i} 
+ k^\gamma c_0 - p' b_0\right) |v, u; k \rangle_R \ . \label{nrDiraceq1} 
\end{eqnarray}
The first equation is the usual mass shell condition. The second equation is an analogue of the 
relativistic Dirac equation. We can check that $G_0^2 = L_0$. So the $G_0$ condition implies the 
mass shell condition.   

The second equation is particularly important for us to investigate the difference between the spectrum 
of the non-relativistic theory and that of the relativistic one. To make things more transparent, 
we can rewrite the equation in terms of the fields $\psi^0$ and $\psi^1$, which reads 
\begin{eqnarray}
\frac{1}{2^{1/2}} \Big( \alpha'^{1/2} k^i \psi_{0, i}  
- ( k^\gamma + p' ) \psi_{0, 0} + (k^\gamma - p') \psi_{0, 1} \Big) = 0 \ .
\label{nrDiraceq2}
\end{eqnarray} 
This equation is the same as the relativistic one if we use 
$\left(\frac{\alpha'}{2}\right)^{1/2} k^\mu \psi_{0, \mu} = 0$, with $(\alpha' )^{1/2} k^0 = - k^\gamma - p' $ 
and $(\alpha' )^{1/2} k^1 =  k^\gamma - p' $. With an appropriate signature, we can get 
\begin{equation}
k^\mu k_\mu = \frac{\alpha' ~ k^i k_i}{2} - \frac{(k^\gamma + p')^2 }{2} 
+ \frac{(k^\gamma - p')^2 }{2} =  \frac{\alpha'}{2} k^i k_i - 2 k^\gamma p' = 0 \ .
\end{equation} 
Particularly there is no further constraint in the vertex operators for the change of fields from $bc$ to 
$\psi^0, \psi^1$, thus the fermionic sector has $SO(1,1) \times SO(8)$ symmetry,\footnote{
It is interesting to observe that the one parameter family of superconformal symmetry ``$SO(1,1)$'' can be 
transformed into $SO(1,1)$ Lorentz symmetry.} 
where there is no 
connection between $\psi^0, \psi^1$ and the other $\psi^i$s. 
It is interesting to observe that the $SO(1,1)$ has boost symmetry and is realized 
as the rescaling of the relative magnitude of $k^\gamma$ and 
$p'$ while keeping the magnitude of their product $k^\gamma p'$ fixed.

We can think about the non-relativistic Dirac equation with manifest $SO(8)$ symmetry structure. 
For the spinors of $SO(8)$, we can impose the Majorana condition and the Weyl condition simultaneously, 
and there are two inequivalent irreducible spinor representations, ${\bf 8}_c$ and ${\bf 8}_s$.
The description of Dirac matrices for $SO(8)$ requires a Clifford algebra with eight anticommuting matrices, 
which are 16-dimensional matrices corresponding to reducible ${\bf 8}_c + {\bf 8}_s$ representation of $SO(8)$. 
These matrices can be written in the block form 
\begin{equation}
\gamma^i = \begin{pmatrix} 
0 & \gamma^i_{a\dot{a}} \\ \gamma^i_{\dot{b}b} & 0 
\end{pmatrix} \ ,
\end{equation}
where the equations $\{\gamma^i , \gamma^j \} = 2 \delta^{ij}$ are satisfied with 
$\gamma^i_{a\dot{a}} \gamma^j_{\dot{a}b} + \gamma^j_{a\dot{a}} \gamma^i_{\dot{a}b} = 2\delta^{ij} \delta_{ab}$ 
with $i,j = 2, \cdots , 9$. $\gamma^i_{\dot{a}a}$ is the transpose of $\gamma^i_{a\dot{a}}$ and can be 
expressed in terms of real components.

To apply these matrices to the non-relativistic Dirac equation (\ref{nrDiraceq2}), we can construct the ten 
dimensional Dirac matrices $\Gamma^\mu$ explicitly 
\begin{equation}
\Gamma^0 = \sigma^3 \otimes {\bf 1}_{16} ,\qquad \Gamma^1 = \sigma^1 \otimes {\bf 1}_{16}, \qquad
\Gamma^0 = i\sigma^2 \otimes \gamma^i,
\end{equation}
where ${\bf 1}_{16}$ is the $16 \times 16$ identity matrix and $i = 2, \cdots , 9$. Here all the 
Gamma matrices are real and thus it is possible to impose Majorana condition for all the spinor fields. 
Using $\psi_0^\mu = \Gamma^\mu/\sqrt{2}$, we can rewrite the equation (\ref{nrDiraceq2}) as 
$\frac{\alpha'^{1/2}}{2} ~k_\mu \Gamma^\mu = 0$. To go further we can use the basis 
\begin{equation}
v_{s_0} \otimes u_{\vec{s}} = \begin{pmatrix} v_+ \\ v_- \end{pmatrix}_2 \otimes 
\begin{pmatrix} u^b \\ u^{\dot{a}} \end{pmatrix}_{16} \ .
\end{equation}
And we can explicitly write the non-relativistic Dirac equation 
\begin{equation}
\frac{\sqrt{\alpha'}}{2} \begin{pmatrix} v_+ \\ -v_- \end{pmatrix}_2 \otimes 
\begin{pmatrix} k_i \gamma^i_{a\dot{a}}u^{\dot{a}} \\k_i \gamma^i_{\dot{b}b} u^{b} \end{pmatrix}_{16}
+ \begin{pmatrix} k^\gamma v_- \\ -p' v_+ \end{pmatrix}_2 \otimes 
\begin{pmatrix} u_a \\ u_{\dot{b}} \end{pmatrix}_{16} = 0 \ .
\end{equation}
To solve this problem we can go to a basis $v_+ = \sqrt{\frac{k^\gamma}{p'}}~ v_-$.\footnote{ 
This condition is actually equivalent to use the symmetry transformation of $SO(1,1)$ to rescale $k^\gamma = p'$. 
}
Then we have the equations,
\begin{eqnarray}
\frac{\sqrt{\alpha'}}{2} k_i \gamma^i_{a\dot{a}} u^{\dot{a}} + \sqrt{k^\gamma p'} u_a &=& 0 \ , \\
\frac{\sqrt{\alpha'}}{2} k_i \gamma^i_{\dot{b}b} u^{b} + \sqrt{k^\gamma p'} u_{\dot{b}} &=& 0 \ .
\end{eqnarray}
These equations are very similar to the relativistic Dirac equation presented in \cite{gsw} with a definite 
chirality in the 10 dimensional fermion.\footnote{We thank Professor Petr Ho\v{r}va for discussions and comments 
on the non-relativistic Dirac equation and interesting ideas related to non-relativistic system.} 
And it is possible to satisfy the non-relativistic 
Dirac equation with manifest $SO(8)$ symmetry by exploiting the superconformal rescaling symmetry. Furthermore 
this equation tells that there is no chiral property for the non-relativistic fermions because these two 
inequivalent irreducible spinor representations ${\bf 8}_c$ and ${\bf 8}_s$ are connected by the non-relativistic 
Dirac equation.\footnote{
Then why are there two inequivalent propagating degrees of freedom ${\bf 8}_s$ and ${\bf 8}_c$ 
in the relativistic case? 
These two inequivalent degrees of freedom come from the 10 dimensional Weyl conditions 
$\Gamma_{11} \lambda = \pm \lambda$, which are not available for the non-relativistic theory. 
For $k^0 = k^9$, it is possible to impose $s_0 = 1/2$ and there is ${\bf 8}_s$ spinor. For $k^0 = -k^9$, the other 
spinor ${\bf 8}_c$ is available. (These two equations $k^0 = \pm k^9$ satisfy $k^\mu k_\mu = 0$.) 
This does not apply for the non-relativistic theory. Because there is no 10 dimensional Weyl condition and 
the bosonic dispersion relation does not have two inequivalent choice for the relation $k^\gamma$ and $p'$.
} 
We will denote this as ${\bf 8}$. Thus we can summarize the particle contents for the first two states in 
the $NS$ sector and for the ground state of $R$ sector in the table.

\bigskip
\begin{table}[h]
\begin{tabular}{lcccccc}
\hline \hline
sector  & $\hspace{0.5 in}$ & $SO(8)$ spin & $\hspace{0.5 in}$ & $-\frac{\alpha'}{4} \vec{k}^2 + k^\gamma p'$  \\
\hline
$NS_0$  && ${\bf 1}$ &    & -1/2  \\
$NS$   && ${\bf 8}_v $ && 0   \\
$R$    && ${\bf 8}$  & & 0  \\
\hline \hline
\end{tabular}
\caption{Spectrum of the holomorphic sector for ground and first excited level of $NS$ sector and 
ground state of $R$ sector. ${\bf 8}_v$ is the fundamental representation of $SO(8)$ 
and ${\bf 8}$ is one copy of the spinor representation of $SO(8)$. }
\end{table}
\bigskip

{\it Closed String Spectrum}

The closed string spectrum has two copies of above spectrum, each from holomorphic and antiholomorphic sectors. 
Because of the level matching condition $NS_0$ sector can only combined with the other $NS_0$ sector 
$-\frac{\alpha'}{4} \vec{k}^2 + k^\gamma p' = -\frac{\alpha'}{4} \vec{k}^2 + k^{\bar{\gamma}} q' = -1/2$. 
This is a nondegenerate state of the non-relativistic closed string. This state will be projected out 
due to the requirement of modular invariance which requires at least one $R$ sector. 

Now it is rather straightforward to construct the closed string spectrum at the next level because there are 
one copy of the vector representation ${\bf 8}_v$ and one copy of the spinor representation ${\bf 8}$ of $SO(8)$. 
The spinor representation ${\bf 8}$ is nonchiral and it is expected that the whole theory is nonchiral. 
We can identify the spinor representation ${\bf 8}$ as one of the two chiral representations 
${\bf 8}_c$ or ${\bf 8}_s$ of $SO(8)$.
And then the whole spectrum is similar to that of the relativistic Type IIB superstring theory, which 
has the same spinor representations in both the holomorphic and the antiholomorphic sectors. 
This signals that the theory is modular invariant and consistent even before we actually check the modular 
invariance. We summarize the ground state and first excited states in the following table. 

\bigskip
\begin{table}[h]
\begin{tabular}{lcccccc}
\hline \hline
sector  & $\hspace{0.25 in}$ & $SO(8)$ spin & $\hspace{0.5 in}$ & tensors & $\hspace{0.5 in}$ & dimensions \\
\hline 
($NS_0$, $NS_0$) && ${\bf 1} \times {\bf 1}$ && &=&  ${\bf 1}$  \\ 
\hline
($NS$, $NS$) && ${\bf 8}_v \times {\bf 8}_v$ &= & [0] + [2] + (2) &=&${\bf 1} + {\bf 28} + {\bf 35}$ \\
($NS$, $R$) && ${\bf 8}_v \times {\bf 8}$ & &&=& ${\bf 8} + {\bf 56}$  \\
($R$, $NS$) && ${\bf 8} \times {\bf 8}_v$ & &&=& ${\bf 8} + {\bf 56}$  \\
($R$, $R$) && ${\bf 8} \times {\bf 8}$ &=& [0] + [2] + [4] &=& ${\bf 1} + {\bf 28} + {\bf 35}$  \\
\hline \hline
\end{tabular}
\caption{Closed superstring spectrum for the ground state and the first excited state of $NS$ sector and 
the ground state of $R$ sector. ${\bf 8}_v$ is the fundamental representation and ${\bf 8}$ is one copy of 
the spinor representation of $SO(8)$. }
\end{table}
\bigskip

\subsection{Partition Function and Modular Invariance}

To show that the theory is consistent, we need to check the modular invariance. 
The bosonic part of the modular invariance is already shown in the previous work\cite{Kim:2007hb}. 
Thus we can concentrate on the fermionic sector. As explained in the previous section, 
The field contents of the non-relativistic superstring theory is the same as those of the relativistic IIB string 
theory. Thus the modular invariance can be proved in a similar way.
For completeness we provide a very brief proof of the modular invariance of the fermionic sector 
by closely following \cite{pol}.

For the complex fermion $\psi$, we can introduce a general periodicity $\alpha = 1- 2\nu$ with 
\begin{equation}
\psi (\omega + 2\pi) = e^{\pi i (1-\alpha)} \psi (\omega) \ .
\end{equation} 
Then the raising operators can be written as $\psi_{-m + (1-\alpha)/2}$ and 
$\psi^{\dagger}_{-m + (1+\alpha)/2}$ with positive integer $m$.
In the bosonized language given in (\ref{bosonizedvertex}), the weight of the vertex operator is  $\alpha^2/8$.\footnote{
We can get the same result from the fermionic language, where the normal ordering constant can be calculated by 
the zero point mnemonic given in \cite{pol}. } 
Using this result we can calculate 
\begin{equation}
Tr_\alpha \Big( q^{L_0 - c/24} \Big) = q^{(3 \alpha^2 - 1)/24} \prod_{m=1}^{\infty} 
\left( 1 + q^{m - (1-\alpha)/2} \right) \left(1 + q^{m - (1+\alpha)/2}\right) \ .
\end{equation}
To accommodate this general boundary condition, we join the fermions into complex pairs in \ref{complexfermions11}. 
Then a fermion number $Q$ can be 
defined as $+1$ for $\psi$ and $-1$ for $\psi^{\dagger}$. $Q$ corresponds to be $H$ momentum in the bosonization 
formula. The ground state has a $Q$ charge as $\alpha/2$. Thus we can define the more general trace 
\begin{eqnarray}
Z_\beta^\alpha (\tau) &=& Tr_\alpha \Big( q^{L_0 - c/24} \exp(\pi i \beta Q) \Big) 
= q^{(3 \alpha^2 - 1)/24} \exp(\pi i \alpha \beta /2)  \\ 
&& \times\prod_{m=1}^{\infty}
\left( 1 + \exp(\pi i \beta) q^{m - (1-\alpha)/2} \right) \left(1 + \exp(-\pi i \beta) q^{m - (1+\alpha)/2}\right) \\
&=& \frac{1}{\eta(\tau)} \vartheta \left[ \begin{array}{cl} \alpha/2 \\ \beta/2 \end{array} \right](0, \tau) \ .
\end{eqnarray}
Here $\alpha$ and $\beta$ can have $0$ and $1$. We have the relevant traces $Z_0^0, Z_0^1, Z_1^0$ and $Z_1^1$. 
The holomorphic part of the partition function for the fermionic sector is 
\begin{equation}
Z_{\psi}(\tau) = \frac{1}{2} \Big[ Z_0^0 (\tau)^4 - Z_1^0(\tau)^4 - Z_0^1 (\tau)^4 - Z_1^1 (\tau)^4 \Big] \ ,
\end{equation}
where the first $-$ sign comes from the ghost contribution and the last two $-$ signs come from 
the spacetime spin statistics. And the total partition function is 
\begin{eqnarray}
Z_{total} = \frac{V_{8} V_{\beta\gamma}}{2p'q'} \int_F \frac{d^2 \tau}{16 \pi^2 \alpha' \tau_2^2}
\left(Z_X^{8} Z_{\psi}(\tau) Z_{\psi}(\tau)^*\right) \ .
\end{eqnarray}
This short explanation proves the modular invariance and it is the same as that of the Type IIB string.

\section{General Non-Relativistic Supersymmetric String}

In this section we consider the $\beta\gamma$ and $bc$ CFTs with general conformal weights. 
First we explain the new matter sector in the superspace formulation. Then we construct a ``noncritical'' version of 
the non-relativistic superstring theories.

\subsection{Matter ${\bf \Sigma\Gamma}$ CFT}

\noindent Let's start with supersymmetric string theory action with a matter ${\bf \Sigma\Gamma}$ 
CFT in addition to the usual ${\bf X}^i$ CFT and the ghost ${\bf BC}$ CFT in the conformal gauge 
\begin{equation}
S_{susy} = \int \frac{d^2 z d^2 \theta}{2\pi} \left( {\bf \Sigma} \bar{\bf D}_{\bar{\theta}} {\bf \Gamma} \right). 
\label{susyaction1} 
\end{equation}
The equations of motion for the fields are $\bar{\bf D}_{\bar{\theta}} {\bf \Gamma} = 0 
= \bar{\bf D}_{\bar{\theta}} {\bf \Sigma} .$
There are a similar action and equations of motion for the anti-holomorphic part of ${\bf \Sigma\Gamma}$ and ${\bf BC}$ CFTs. 

OPEs of new ${\bf \Sigma\Gamma}$ CFT are given by 
\begin{equation}
{\bf \Gamma}(z_1, \theta_1) {\bf \Sigma}(z_2, \theta_2) ~\sim~ 
\frac{\theta_{12}}{ \hat{z}_{12}} ~\sim~ {\bf \Sigma}(z_1, \theta_1) {\bf \Gamma}(z_2, \theta_2) \ ,
\end{equation}
where $\theta_{12} = \theta_1 - \theta_2$ and $\hat{z}_{12} = z_1 - z_2 -\theta_1 \theta_2$. 
The super energy momentum tensor\footnote
{This can be contrasted to the energy momentum tensor of ${\bf BC}$ super ghost CFT 
\begin{equation}
{\bf T}_{ghost}^{{\bf B}{\bf C}} = - (\lambda_g - 1) {\bf C}\left({\bf D}^2 {\bf B} \right) 
+ \frac{1}{2} \left({\bf D} {\bf C}\right) \left({\bf D} {\bf B} \right) 
- (\lambda_g - \frac{1}{2}) \left({\bf D}^2 {\bf C}\right) {\bf B} \ .  \nonumber
\end{equation}
The ghost energy momentum tensor has the same form as that of the matter ${\bf \Sigma\Gamma}$ CFT except the sign differences. 
And the conformal weights of the ghost super fields with $\lambda_g = 2$ are $h({\bf B}) = \lambda_g - 1/2,
h({\bf C}) = 1 - \lambda_g$. And those of the component fields are $ h(\beta_g) = \lambda_g - 1/2, h(c_g) 
= 1 - \lambda_g, h(b_g) = \lambda_g, h(\gamma_g) = 3/2 - \lambda_g.$}
is a chiral superfield of dimension $3/2$ with the ordinary 
energy momentum tensor of dimension $2$ in it ${\bf T}({\bf z}) = T_F(z) + \theta T_B(z)$ 
\begin{equation}
{\bf T} = (\lambda - 1) {\bf\Gamma} \partial {\bf \Sigma} 
+ \frac{1}{2} \left( {\bf D} {\bf \Gamma}\right) \left( {\bf D} {\bf \Sigma} \right) 
+ (\lambda - \frac{1}{2}) \partial {\bf \Gamma} {\bf \Sigma} \ .
\end{equation}
For $\lambda = 1$ case, the super energy momentum tensor simplifies further and have the form 
\begin{equation}
{\bf T}_{\lambda = 1} = \frac{1}{2} \left( {\bf D} {\bf \Gamma}\right) \left( {\bf D} {\bf \Sigma} \right) 
+ \frac{1}{2} \partial {\bf \Gamma} {\bf \Sigma} \ ,
\end{equation}
which is very simple and we concentrate on the previous section as a critical case. It is simple to verify that this reduces to  
the component forms of the energy momentum tensor (\ref{bosonicenergymomentumtensor111}) and (\ref{bosonicenergymomentumtensor222}), 
which are presented below. 
The case with $\lambda = 1/2$ also simplifies and corresponds to the ``critical'' case in a sense we explain in 
the next subsection. 

The super energy momentum tensor is itself an anomalous superconformal field 
\begin{equation}
{\bf T}(z_1, \theta_1) ~{\bf T}(z_2, \theta_2) \sim \frac{8\lambda - 6}{4 \hat{z}_{12}^3} 
+ \frac{3}{2} \frac{\theta_{12}}{\hat{z}_{12}} {\bf T}(z_2, \theta_2) 
+ \frac{1}{2}\frac{1}{\hat{z}_{12}} {\bf D}_2 {\bf T}(z_2, \theta_2) 
+ \frac{\theta_{12}}{z_{12}} \partial_2 {\bf T}(z_2, \theta_2) \ ,
\end{equation}
which tells us the central charge of super energy momentum tensor is 
$\hat{c} = \frac{2}{3} c = 8\lambda - 6$ and the conformal weight of the tensor is $3/2$. 

OPEs of the energy momentum tensor with the super fields can be calculated 
\begin{eqnarray}
{\bf T}(z_1, \theta_1) ~{\bf \Gamma}(z_2, \theta_2) &\sim& 
(1 - \lambda) \frac{\theta_{12}}{\hat{z}_{12}^2} {\bf \Gamma} (z_2, \theta_2) 
+ \frac{1}{2}\frac{1}{\hat{z}_{12}} {\bf D}_2 {\bf \Gamma} (z_2, \theta_2) 
+ \frac{\theta_{12}}{\hat{z}_{12}} \partial_2 {\bf \Gamma} (z_2, \theta_2) \ , \nonumber \\
{\bf T}(z_1, \theta_1) ~{\bf \Sigma}(z_2, \theta_2) &\sim& (\lambda - \frac{1}{2}) 
\frac{\theta_{12}}{\hat{z}_{12}^2} {\bf \Sigma} (z_2, \theta_2) 
+ \frac{1}{2}\frac{1}{\hat{z}_{12}} {\bf D}_2 {\bf \Sigma} (z_2, \theta_2) 
+ \frac{\theta_{12}}{\hat{z}_{12}} \partial_2 {\bf \Sigma} (z_2, \theta_2) \ .
\end{eqnarray}
These equations tells us that the new fields ${\bf \Gamma}$ and ${\bf \Sigma}$ have conformal weights 
$h({\bf \Gamma}) = 1 - \lambda$ and $h({\bf \Sigma}) = \lambda - 1/2$, respectively. 

The dimensions of the component fields are 
\begin{eqnarray}
{\bf \Gamma} = -\gamma + \theta c \ , & h(\gamma) = 1 - \lambda  \ , & h(c) = 3/2 - \lambda  \ , \\
{\bf \Sigma} = b + \theta \beta \ , & h(b) = \lambda -1/2 \ , & h(\beta) = \lambda  \ .
\end{eqnarray}
And $\gamma, \beta$ and ${\bf \Gamma}$ are commuting fields and $b, c$ and ${\bf \Sigma}$ are anticommuting fields. 

Using the component fields we can rewrite the supersymmetric action
\begin{eqnarray}
S_1 &=&  \int \frac{d^2z}{2\pi} \left( \beta \bar{\partial} \gamma + \bar{\beta} \partial \bar{\gamma} 
+ b \bar{\partial} c + \bar{b} \partial \bar{c} \right) \ .
\end{eqnarray}
Given the conformal weights of the component fields, the central charge of the $\beta\gamma$ CFT and the $bc$ CFT are 
$3(2\lambda - 1)^2 -1$ and $-3(2\lambda -2)^2 + 1$, respectively. Thus the total central charge is $c = 12 \lambda - 9$, 
which agrees with the result from the OPE of the energy momentum tensor.   

And the OPEs of the component fields are  
\begin{eqnarray}
\gamma(z_1) \beta(z_2) \hspace{0.1 in}  \sim & \frac{1}{z_{12}} & \sim  \hspace{0.1 in} 
 - \beta(z_1) \gamma(z_2) \ , \\
b(z_1) c(z_2)  \hspace{0.1 in}  \sim & \frac{1}{z_{12}} & \sim  \hspace{0.1 in}  c(z_1) b(z_2) \ .
\end{eqnarray} 
The energy momentum tensor in the component form can be written 
\begin{eqnarray}
T_{b} = (\lambda -\frac{3}{2}) c (\partial b) 
+ (\lambda -\frac{1}{2})(\partial c) b
- (\lambda - 1) \gamma (\partial \beta) - \lambda (\partial \gamma) \beta 
&=& \sum_{m \in {\bf Z}} \frac{L_m}{z^{m+2}} \ , \\
T_{f} = -(\lambda - 1) \gamma (\partial b) + \frac{1}{2} c \beta 
- (\lambda - \frac{1}{2}) (\partial \gamma) b 
&=& \sum_{r \in {\bf Z} + \nu} \frac{G_r}{2 \cdot z^{r+3/2}} \ . 
\end{eqnarray}

As is well known, the fields with the half integer conformal weight have both $NS$ and $R$ sectors. 
To make the expressions simple, we concentrate on the case of integer $\lambda$. 
The mode expansions and the hermiticity properties are 
\begin{eqnarray}
\gamma(z) = \sum_{n \in {\bf Z}} \frac{\gamma_n}{z^{n + 1-\lambda}} \ , \hspace{0.3 in} \gamma_n^{\dagger} =  \gamma_{-n} \ , 
&& \beta(z) = \sum_{n \in {\bf Z}} \frac{\beta_n}{z^{n+ \lambda}} \ , \hspace{0.3 in} \beta_{n}^{\dagger} =  - \beta_{-n} \ ,\\
c(z) = \sum_{r \in {\bf Z}+ \nu} \frac{c_r}{z^{r+3/2-\lambda}} \ , \hspace{0.2 in} c_r^{\dagger} =  c_{-r} \ ,
&& b(z) = \sum_{r \in {\bf Z} + \nu} \frac{b_r}{z^{r+\lambda- 1/2}} \ , \hspace{0.2 in} b_{r}^{\dagger} =  b_{-r} \ .
\end{eqnarray}
There are two possible values for $\nu$. For the $NS$ sector $\nu = 1/2$ and for $R$ sector $\nu = 0$. 
And the mode expansions for the energy momentum tensors are 
\begin{eqnarray}
&& L_m^{\beta\gamma b c} = \sum_{n \in {\bf Z}}\Big( n - (1-\lambda)m \Big) \beta_{m-n}\gamma_n 
- \sum_{s \in {\bf Z} +\nu} \Big(s - (3/2 - \lambda )m \Big) b_{m-s} c_s 
+ a\delta_{m,0} \ , \\
&& G_r^{\beta\gamma b c} = \sum_{n \in {\bf Z}} \left( c_{r-n} \beta_n + \Big( n + 2 r (\lambda - 1)\Big) \gamma_n b_{r-n} \right) \ .
\end{eqnarray}
There is a normal ordering constant in each sector, $ a_R^{\beta\gamma b c} = \frac{4\lambda - 3}{8}$ and 
$a_{NS}^{\beta\gamma b c} = \frac{\lambda - 1}{2}$.

\subsection{Possible Non-Relativistic Superstring Theories}

\noindent It is interesting to construct a ``noncritical'' version of the non-relativistic superstring theory.
Central charge of the ghost part is $\hat{c}_{{\bf BC}} = -10$ and that of the matter 
CFT is $\hat{c}_{\bf \Sigma\Gamma} = 8\lambda - 6$. Thus to be consistent the dimension $D$ 
of the spatial directions in target space is 
\begin{equation}
D = 8(2 - \lambda).
\end{equation}
We summarized the interesting portion of theories in the table. 

\bigskip
\begin{table}[h]
\begin{tabular}{c||c|c|c|c|c|c|c|c|c}
\hline \hline
$\lambda$
  & \hspace{0.1 in}$\cdots$\hspace{0.1 in} & \hspace{0.1 in}2\hspace{0.1 in} &\hspace{0.1 in} $\frac{3}{2}$\hspace{0.1 in} 
&\hspace{0.1 in}1\hspace{0.1 in} & \hspace{0.1 in}$\frac{1}{2}$ \hspace{0.1 in} &\hspace{0.1 in} 0\hspace{0.1 in} 
&\hspace{0.1 in}- $\frac{1}{2}$\hspace{0.1 in} &\hspace{0.1 in} -1\hspace{0.1 in} & \hspace{0.1 in}$\cdots$\hspace{0.1 in} \\
\hline
$\hat{c}_{\bf \Sigma\Gamma} = 8\lambda - 6$
  & $\cdots$ & 10 & 6  & 2 & -2 & -6 & -10 & -14 & $\cdots$ \\
\hline
$D = 8(2 - \lambda)$
 & $\cdots$ & 0 & 4 & 8 & 12 & 16 & 20 & 24 & $\cdots$   \\
\hline \hline
\end{tabular}
\caption{Table for the super string case. Conformal weight 
of the supersymmetric $\beta\gamma$ CFT and the number of spatial dimensions 
of target space are presented. For $\lambda > 2$, the geometric interpretation is not possible. 
As the parameter $\lambda$ is decreasing, the number of spatial dimensions is growing indefinitely and linearly. 
}
\label{table:fermionictable3}
\end{table}
\bigskip

Here we comment on the immediate observations of these possible consistent ``noncritical'' 
non-relativistic superstring theories. These theories have the same actions and the $SO(1,1) \times SO(D)$ 
symmetries in addition to Galilean symmetry. There exists an infinite range of possible consistent theories with 
geometric interpretation, for which we mean it is possible to have positive number of spatial coordinates. 

It will be interesting to quantize them explicitly. We can divide them in two categories, 
(i) with integer $\lambda$ cases and (ii)
with half integer $\lambda$ cases, because there are two sectors for the fields with half integer 
conformal weight. For the integer $\lambda$ cases (i) with $D = 0, 8, 16, \cdots$, 
the bosonic commuting $\beta\gamma$ CFT has only one bosonic coordinate. 
From the explicit quantization of the previous section and from \cite{Kim:2007hb}, 
we know that it is relatively easy to quantize and establish the spacetime interpretation. 
On the other hand, there are two commuting bosonic sectors, $NS$ and $R$, for the half integer $\lambda$ 
cases (ii) with $D = 4, 12, 20, \cdots $. Of course, in case (ii) the 
zero modes of the $R$ sector of the $\beta\gamma$ CFT have a space and time interpretation. 
The case (ii) seems rather peculiar and it looks harder to quantize them. But these theories are expected 
to provide different perspective for a space and time interpretation. 

The challenges of establishing the zero modes of $\beta\gamma$ CFT in the new matter sector 
can be easily seen by the total normal ordering constant. As usual, the normal ordering 
constant for the $R$ sectors is $0$ due to the cancellation between the bosonic contribution and the fermionic 
contribution. And those of the $NS$ sectors are 
\begin{eqnarray}
a^{(i)}_{NS} = \frac{\lambda -2}{2} \ ,  &  \hspace{0.4 in} &  a^{(ii)}_{NS} = \frac{2\lambda -3}{4} \ .
\end{eqnarray}
Thus the total normal ordering constant for the $NS$ sector depends on the parameter 
$\lambda$ and there is nontrivial mapping between the unit vertex operator $1$ and the 
corresponding state. We can see that the case with $\lambda = 1$, we considered in the previous section, 
is critical in the sense that the normal ordering constants $a^{(i)}_{NS} = - \frac{1}{2}$ recover 
those of the critical relativistic string theory. It is interesting to comment that there is another 
``critical'' case for the case (ii) with $\lambda = \frac{1}{2}$. Thus the cases with $\lambda = 1$ and 
$\lambda = \frac{1}{2}$ tie together in a sense and we expect that the space and time interpretation is rather 
similar. This observation extends to all the other cases. The case with $\lambda = n $ and 
$\lambda = n + \frac{1}{2}$ tie together for integer $n$. Quantization of the theory with $\lambda = \frac{1}{2}$ 
and comparison to the critical case with $\lambda = 2$ will be very interesting. 

In the case $\lambda = 2$ with $D=0$, there are only ${\bf \Sigma \Gamma}$ 
CFT and ${\bf BC}$ CFT. Upon quantization, only the zero modes are present without oscillator excitations. 
The theory is topological. Furthermore there is a possible unification of these CFTs in a simple fashion. 
We comment this at the end of this section. As explained in the previous paragraph, this case is tied with 
the $\lambda = \frac{3}{2}$ case in a sense that the normal ordering constant is same and thus the zero modes 
have similar roles. But this is not a ``topological'' case because there are additional 4 spatial coordinates. 

\bigskip
\noindent{\it unification of all the first order CFTs } 
\smallskip

There is a curiosity related to a possible interesting ${\bf Z}_2$ graded algebra involving the 
nonzero conformal weight, the U(1) ghost number and the U(1) number of the matter ${\bf \Sigma\Gamma}$ CFT. We can make 
a table for basic properties of the first order matter CFT and the ghost CFT 

\bigskip
\begin{table}[h]
\begin{tabular}{|l|c|c|c||c||c|c|c|c|}
\hline
field & weight & $ U(1)^{\bf m}$ & $U(1)^{\bf gh}$ & \hspace{0.5 in} & field & weight 
&$ U(1)^{\bf m}$ &$ U(1)^{\bf gh}$ \\
\hline \hline
$b_g$ & $\lambda_g$ & 0 & $-1$ & & $c_g$ & $1-\lambda_g$ & 0 & 1  \\
\hline
$\beta_g$ & $\lambda_g - 1/2$ & 0 & $-1$ & & $\gamma_g$ & $3/2-\lambda_g$ & 0 & 1  \\
\hline
$\beta$ & $\lambda$ & $-$1 & $0$ & & $\gamma$ & $1-\lambda$ & 1 & 0  \\
\hline
$b$ & $\lambda -1/2$ & $-1$ & 0 & & $c$ & $3/2-\lambda$ & 1 & 0  \\
\hline
\end{tabular}
\caption{Table for the various properties of the first order matter CFT and the ghost CFT. 
We list the conformal weight, U(1) charge of the matter $\beta\gamma$ CFT and U(1) charge 
of the ghost CFT.
}
\label{table:fermionictable4}
\end{table}

From this table we can imagine that there are two grand supermultiplets 
${\bf V}$ and ${\bf W}$ with new field $\Theta_{gh}$ which carries conformal weight, 
U(1) ghost charge and U(1) matter charge  
\begin{eqnarray}
{\bf V} = {\bf \Sigma} + \Theta_{gh} {\bf B} = b + \theta \beta 
+ \Theta_{gh} (\beta_g + \theta b_g)= b + \Theta_{gh} \beta_g + \theta (\beta 
+ \Theta_{gh} b_g), \\
{\bf W} = {\bf C} + \Theta_{gh} {\bf \Gamma} = c_g 
+ \theta \gamma_g + \Theta_{gh} (-\gamma + \theta c) = c_g - \Theta_{gh} \gamma + 
\theta (\gamma_g + \Theta_{gh} c).
\end{eqnarray}
If one investigates these grand multiplets a little further one can read off 
that $\Theta_{gh}$ is anticommuting field with conformal weight 
$\lambda - \lambda_g$, matter U(1) charge $-1$ and ghost number $1$. 
${\bf V}$ is an anticommuting multiplet with the conformal weight $\lambda -1/2$, 
the U(1) matter charge $-1$ and the ghost U(1) number $0$, 
whereas ${\bf W}$ is an anticommuting multiplet with the conformal weight $1-\lambda_g $, 
the U(1) matter charge $0$ and the ghost U(1) number $1$.  
We comment on two cases with immediate interest. 
One is $\lambda = 1$ case with the conformal weight of the field $\Theta_{gh}$ as $-1$.
Then all the fields have uniform gaps of their conformal weights. This is the case we quantized in the 
previous section. For $\lambda = 2$, the field $\Theta_{gh}$ has no conformal weight. 
This is a topological case with these two multiplets only without other matter sector.

With these observation we can rewrite the superstring action in a very simple form 
for holomorphic part
\begin{equation}
S_{{\bf V}{\bf W}} = \int \frac{d^2 z d^2 \theta}{2\pi} d\Theta_{gh} \Big({\bf V} 
\bar{\bf D}_{\bar{\theta}} {\bf W}  \Big) = \int \frac{d^2 z d^2 \theta}{2\pi}
(\Sigma \bar{\bf D}_{\bar{\theta}} {\bf \Gamma} + \
{\bf B}\bar{\bf D}_{\bar{\theta}} {\bf C} ) \label{vwaction} 
\end{equation}
Note that this action has still the derivative of the form 
$\bar{\bf D}_{\bar{\theta}} = \partial_{\bar{\theta}} + \bar{\theta} \partial_{\bar{z}}$ and we did not 
gauge the field $\Theta_{gh}$. It will be interesting if we can gauge the field $\Theta_{gh}$.

\section{Non-Relativistic Strings with Higher Supersymmetry}

Following Polchinski \cite{pol}, we would like to survey possible superconformal algebras 
and their related non-relativistic superstring theories. The basic idea is to find the sets of holomorphic 
and antiholomorphic currents, whose Laurent coefficients form a closed constraint algebra. This is 
motivated by the idea of enlarging the world sheet constraint algebra with supercurrents $T_F(z)$ and 
$\bar{T}_F(\bar{z})$. Here the constraint is part of the symmetry singled out to be imposed on physical 
states in OCQ or BRST sense. 

Here we assume that there is only one $(2,0)$ constraint current because the sum of the $\beta\gamma$, 
$bc$ and $X^i$ energy momentum tensors have geometric interpretation in terms of conformal invariance. 
This is similar to the relativistic case. Thus the result of the constraint current algebra in world sheet 
is the same as the relativistic case.  
Concentrating on holomorphic current with conformal weight as multiple of half integer and less than and 
equal to 2,\footnote{
For the ghost CFT, there are restrictions as we mentioned. But there is no restriction for the matter 
$\beta\gamma$ or $bc$ CFT because they are part of the $(2,0)$ constraint current and they are consistent 
part of the algebra as long as all the matter conformal weight sums up to satisfy the physical state conditions.
}
there are very limited possible algebras and it is given in the following table.

\begin{table}[h]
\begin{tabular}{c|c|c|c|c|c|c|c|c|c}
\hline \hline
& $n_2$
  & $n_{3/2}$ & $n_1$ & $n_{1/2}$ & $n_0$ & $c_{gh}$ & $c^{\bf m}_{\beta\gamma, bc, \cdots}$ 
& symmetry & $T_F$ Rep. \\
\hline \hline
I&1 & 0 & 0 & 0 & 0 & $-26$ & 
2(6$\lambda^2-6\lambda +1$) &  &  \\
\hline
II&1 & 1 & 0 & 0 & 0 & $-15$ & 
12 $\lambda -9$ &  &  \\
\hline
III&1 & 2 & 1 & 0 & 0 & $-6$ & 
+6 & U(1) & $\pm 1$ \\
\hline
IV&1 & 3 & 3 & 1 & 0 & 0 & 
0 & SU(2)  & {\bf 3} \\
\hline
V&1 & 4 & 7 & 4 & 0 & 0 & 
$24(\lambda -2)$ & $SU(2)^2 \times U(1)$ & ({\bf 2}, {\bf 2}, 0)  \\
\hline
VI&1 & 4 & 6 & 4 & 1 & 0 & 
0 & $SU(2)^2$ & ({\bf 2}, {\bf 2}) \\
\hline
VII&1 & 4 & 3 & 0 & 0 & 12 & 
$36 - 24 \lambda$ & SU(2) & {\bf 2} \\
\hline \hline
\end{tabular}
\caption{Survey of possible string theory. The first five columns represent  
the number of reparametrization currents with corresponding spins as indicated in the subscript of $n_{spin}$. $n_{3/2}$ represent the number of supersymmetry. $c_{gh}$ is the total central charge of the supersymmetrized ghost CFT and $c^{\bf m}_{\beta\gamma, bc, \cdots}$ is the total central charge of the supersymmetrized $\beta\gamma$ CFT. The last two columns represent the symmetry and the representation of the supercharge. }
\label{table:fermionictable5}
\end{table}

The cases I and II are explained already in the bosonic string theory \cite{Kim:2007hb} and in the 
previous section, respectively. These theories are explicitly quantized and have the non-relativistic dispersion relation. 
The cases III, IV and VI are rather different from the other cases because both the supersymmetric 
ghost ${\bf BC}$ CFT and the ${\Sigma\Gamma}$ CFT have the central charges independent of $\lambda$, 
which are same in magnitude with opposite sign. Thus there is no room for the spatial coordinates. 
But it is still possible to have some geometric interpretation from the matter ${\Sigma\Gamma}$ CFTs. 

In addition to the II case, there are two possible cases with infinite number of possible string 
theories, the cases V and VII. Both cases have 4 super charges in world sheet CFT. For case V, 
the central charge of the superconformal ghost CFTs is $0$ and the central charge of the matter 
${\bf \Sigma\Gamma}$ CFTs is $24(\lambda - 2)$. Thus for $\lambda \leq 2$ cases, it is 
possible to have spatial $X$ CFTs. In the last case, VII, the central charge has positive contribution 
from the ghost CFTs. On the other hand, there are negative contribution from the matter ${\Sigma\Gamma}$ 
CFTs. We can make the parameter $\lambda$ large and there is corresponding string theory. It will be 
interesting to quantize these sets of theories.

\section{Conclusions} 

In this paper we construct a supersymmetric version of the recently constructed non-relativistic string theory. 
The non-relativistic superstring theory has a first order ${\bf \Sigma\Gamma}$ SCFT on top of the 
usual eight second order ${\bf X}$ SCFTs. The fermionic sector has an anticommuting matter $bc$ CFT in addition 
to the eight $\psi^i$ fields. The component fields, $b$ and $c$, have the conformal weights $1/2$. 
These can be transformed into the $\psi^0$ and $\psi^1$ fields, and the fermionic action is the same as that of 
the relativistic superstring theory. The symmetry group is $SO(1,1) \times SO(8)$. 

We quantize the theory in an elementary fashion. In addition to the physical state conditions imposed by energy momentum 
tensor, there exist other conditions from the super current. These give us a non-relativistic analogue of the Dirac 
equation in the ground state of the $R$ sector. 
This equation can be solved with the manifest $SO(8)$ symmetry by exploiting $SO(1,1)$ symmetry. The fermionic spectrum 
is non-chiral because the non-relativistic Dirac equation connects the two irreducible spinor representations 
${\bf 8}_c$ and ${\bf 8}_s$ for the $SO(8)$ group. For the closed string spectrum, modular invariance requires 
to project out the ground state in the $NS$ sector. The spectrum of this theory is very similar to that of 
Type IIB superstring theory, except for the chiral property and the energy dispersion relation. 
The one loop consistency check is straightforward and the theory is modular invariant. 

We present a noncritical version of the non-relativistic superstring theories by generalizing the conformal weight of 
the first order ${\bf \Sigma\Gamma}$ SCFT. It turns out that there is an infinite range of possible 
non-relativistic superstring theories. 
We present some immediate observations related to these possible consistent string theories. 
We further survey possible non-relativistic string theories with extended supersymmetry utilizing the world sheet 
constraint algebra. The matter $\beta\gamma$ CFT (and its supersymmetric partners) combined with the $X$ CFT 
(and its partners) form a $(2,0)$ constraint current (and its partners) to have a geometric interpretation. 
Thus the matter first order CFTs are not constrained severely compared to the ghost sector. 
There are three infinite series of possible string theories: two with the four super charges and 
one with the one super charge, which is considered in the present work. It will be interesting to quantize 
these noncritical non-relativistic string theories.

\section{Future Directions} 

Understanding cosmological singularities such as the Big Bang is an interesting and outstanding problem. 
It requires understanding time-dependent backgrounds in string theory, which are very difficult to analyze\cite{Liu:2002ft}. 
Perturbative string theory breaks down in some spacetime regions where the string coupling becomes large. 
One clean example with the lightlike Linear Dilaton theory was recently proposed in 
\cite{Craps:2005wd}.\footnote{There are some direct generalizations of this simple 
solution\cite{Li:2005sz}. We thank Professor Nobuyoshi Ohta for drawing our attention for these solutions.}   
The Dilaton is proportional to a light cone coordinate, $- X^+$, and the theory is defined as an exact CFT 
that describes string propagating in flat spacetime with the string coupling, $g_s = e^{- Q X^+}$. 
Thus the spacetime is free at late times and strongly coupled at early times. At early times, 
there is a true singularity happening at a finite affine parameter, which requires a matrix string
description as explained in \cite{Craps:2005wd}. 
It appears to be necessary to have a complete nonperturbative description of string theory to understand time 
dependent backgrounds in string theory. There is an interesting nonperturbative formulation of noncritical 
M-theory in (2+1) dimensions using the non-relativistic Fermi liquid and its time-dependent 
solutions\cite{Horava:2005tt}. Earlier work with time-dependent background with closed string tachyon condensation 
can be found in (1+1) noncritical string theory\cite{Karczmarek:2004ph}.

On the other hand there are very interesting developments which emphasize the role of perturbative string theory 
in the analysis of time-dependent backgrounds. It is claimed that a certain spacetime singularity can be replaced 
by a tachyon condensate phase within perturbative string theory\cite{McGreevy:2005ci}. And very recent 
papers\cite{Horava:2007yh} argue, using alternative gauge choices to free world sheet gravitino, 
that spacetime decay to nothing in string and M-theory should be addressed 
at weak string coupling, where the nonperturbative instanton instability is expected to turn into a perturbative 
tachyon instability. See also \cite{Hellerman:2007zz}. 
Similar considerations in supercritical string theories can be found in \cite{Aharony:2006ra, Hellerman:2006nx}. 

It turns out that many interesting cosmological solutions have broken Lorentz symmetry. And it is interesting to 
consider these solutions with their manifest global symmetries. Furthermore fundamental issues related to time, 
especially to ``emergent time'', is not clear (see, {\it e.g.,}\cite{Seiberg:2006wf}). 
Thus it is interesting to consider alternative approaches, which can shed light on time-dependent backgrounds 
and on fundamental issues of time.\footnote{An example which motivates a different approach 
for time can be seen in the low energy limits of open string theory with magnetic and electric $NS-NS$ B-field. 
In the appropriate limits, the theory with electric $NS-NS$ B-field is reduced to noncommutative open string theory
while the theory with magnetic $NS-NS$ B-field reduces to the noncommutative Yang-Mills theory. This suggest that 
time is rather different from space. This is motivated to consider non-relativistic string theories in \cite {Kim:2007hb}.} 
Our current work and a previous paper\cite{Kim:2007hb}, 
motivated by earlier works\cite{Gomis:2000bd, Danielsson:2000gi, Danielsson:2000mu}, provide examples for these 
alternative approaches. 

As we saw in the main body, the non-relativistic string theory shares many features with relativistic 
string theory. The difference between these two theories comes from the replacement of the $X^0$ and $X^1$ CFTs 
by $\beta\gamma$ CFT. This effect is minimal because these matter CFTs are part of the $(2,0)$ constraint current, 
which makes a geometric interpretation possible. As a result, the spectrum is very similar to that of 
Type IIB superstring theory. On the other hand, these non-relativistic string theories provide a very different 
perspective on time. Thus these non-relativistic string theories appear 
to be ideal for investigating general issues related to time-dependent backgrounds with broken Lorentz symmetry, 
such as the lightlike Linear Dilaton theory and supercritical string theories.  

We would like to comment on a few preliminary results for the correspondence between the critical non-relativistic string 
theory and the lightlike LDT.\footnote{This correspondence between the non-relativistic string theory and the 
the lightlike Linear Dilaton theory was pointed out by Professor Petr Ho\v{r}ava. We are grateful for his careful and extensive 
suggestions on this correspondence and on various references.}  
These two theories have the same set of global symmetries, which can be checked with the identification 
$X^+ = t$ in the lightlike LDT case. In the lightcone gauge, the spectrum of the lightlike LDT can be checked to be the same as
that of the non-relativistic string theory. These equivalences are enough for us to be serious about 
investigating the exact mapping between these two theories. 
We hope to report these results in the near future.

\section*{Acknowledgments}

It is pleasure to thank Professor Ori Ganor for encouragements, Professor Petr Ho\v{r}ava for introducing 
crucial ideas and references and Professor Ashvin Vishwanath for answering many questions related to 
properties of non-relativistic system. Their careful comments and extensive discussions were critical to 
make this work possible. I also thank Jordan Carlson, Sharon Jue and Stefan Leichenauer for reading and 
commenting the manuscript. This work was supported in part by 
the Center of Theoretical Physics at UC Berkeley,
and in part by the Director, 
Office of Science,
Office of High Energy and Nuclear Physics, 
of the U.S. Department of
Energy under Contract DE-AC02-05CH11231.

\section*{Disclaimer}
This document was prepared as an account of work sponsored by the United States Government. While this document is believed to contain correct information, neither the United States Government nor any agency thereof, nor The Regents of the University of California, nor any of their employees, makes any warranty, express or implied, or assumes any legal responsibility for the accuracy, completeness, or usefulness of any information, apparatus, product, or process disclosed, or represents that its use would not infringe privately owned rights. Reference herein to any specific commercial product, process, or service by its trade name, trademark, manufacturer, or otherwise, does not necessarily constitute or imply its endorsement, recommendation, or favoring by the United States Government or any agency thereof, or The Regents of the University of California. The views and opinions of authors expressed herein do not necessarily state or reflect those of the United States Government or any agency thereof or The Regents of the University of California.

\section*{Appendix: Physical spectrum with $SO(7)$ symmetry}

In this appendix, we consider a relativistic approach to investigate the spectrum of this 
non-relativistic string theory. It is interesting to compare these results with those in the main text. 

We have $SO(1,1) \times SO(8)$ symmetry and we want to analyze the non-relativistic mass shell condition 
(\ref{massshellforns1}) and the non-relativistic Dirac equation (\ref{nrDiraceq2})
\begin{eqnarray}
\frac{\alpha'}{4} \vec{k}^2 - k^\gamma p' = 0 \ , \label{a1} \\
\frac{1}{2^{1/2}} \Big( \alpha'^{1/2} k^i \psi_{0, i} 
- ( k^\gamma + p' ) \psi_{0, 0} + (k^\gamma - p') \psi_{0, 1} \Big) = 0 \ . \label{a2}
\end{eqnarray} 

Rather than breaking the $SO(1,1)$ symmetry, we can go to a frame, $k^i = 0 $ for $i = 2, \cdots , 8$ and $k^9 \neq 0$, 
which preserves the $SO(1,1) \times SO(7)$ symmetry, to solve these two equations (\ref{a1}) and (\ref{a2}). 
From the quantization procedure we know that there are eight 
physical degrees of freedom. There are only the $SO(7)$ manifest symmetry in the first excited level of the $NS$ sector, 
which has a vector representation ${\bf 7}$ of $SO(7)$. Then where is one extra degrees of freedom? 
It is a ``Dilaton'' originated from the conformal rescaling $SO(1,1)$, 
which transforms as a singlet under $SO(7)$. Thus the first excited level has eight degrees of freedom 
which transform as ${\bf 1 + 7}$ under the $SO(7)$ rotation.  

And then we can solve the non-relativistic Dirac equation (\ref{a2}) by using the $SO(1,1)$ 
symmetry by picking particular values of $k^\gamma$ and $p'$. Then the remaining symmetry group 
$SO(1,1) \times SO(7)$ is broken to $SO(7)$. 
The irreducible spinor representation of the $SO(7)$ group is ${\bf 8}$ as is well known. Thus there are actually 
eight independent degrees of freedom in the ground state of the $R$ sector. And it is obvious that there is no 
chance for the fermions to have any chiral property. We present a table for the holomorphic spectrum with $SO(7)$ symmetry.

\begin{table}[h]
\begin{tabular}{lcccccc}
\hline \hline
sector  &  $\hspace{0.5 in}$ &  $SO(7)$ spin  
& $\hspace{0.5 in}$ & $-\frac{\alpha'}{4} \vec{k}^2 + k^\gamma p'$  \\
\hline
$NS_0$  & &  ${\bf 1}$   && -1/2  \\
NS   & & ${\bf 1 + 7}$ && 0   \\
$R$  & & ${\bf 8}$ && 0  \\
\hline \hline
\end{tabular}
\caption{Spectrum of the holomorphic sector for ground and first excited level of $NS$ sector and 
ground state of $R$ sector. ${\bf 7}$ and ${\bf 8}$ are the vector representation and 
the spinor representation of $SO(7)$, respectively. }
\end{table}

It is straightforward to construct the non-relativistic closed superstring spectrum. 
They are presented below. We would like to have a few comments. Compared the approach with the manifest 
$SO(8)$ symmetry, the $SO(7)$ symmetry is not efficient to describe the physical spectrum. 
Furthermore it is not clear that how we can demonstrate the modular invariance at all. 
The field contents are very similar to the relativistic string theory with a circle compactification. 
But in that case there are discrete momentum modes and discrete winding modes in the twisted sector. 
One the other hand, we have just continuous momentum without compact coordinate or twisted sector.

\bigskip
\begin{table}[h]
\begin{tabular}{lcccccc}
\hline \hline
sector  & $\hspace{0.25 in}$ & $SO(7)$ spin & $\hspace{0.5 in}$ & dimensions \\
\hline 
($NS_0$, $NS_0$) && ${\bf 1} \times {\bf 1}$ &= &  ${\bf 1}$  \\ 
\hline
($NS$, $NS$) && $({\bf 1+ 7}) \times ({\bf 1+ 7})$ &= &${\bf 1} + ({\bf 7 + 7}) + ({\bf 1+ 21 + 27})$ \\
($NS$, $R$) && $({\bf 1 + 7}) \times {\bf 8}$ &= & ${\bf 8} + ({\bf 8 + 48})$  \\
($R$, $NS$) && ${\bf 8} \times ({\bf 1 + 7})$ &= & ${\bf 8} + ({\bf 8 + 48})$  \\
($R$, $R$) && ${\bf 8} \times {\bf 8}$ &= & ${\bf 1} + ({\bf 7 + 21}) + ( {\bf 1 + 7 + 27})$  \\
\hline \hline
\end{tabular}
\caption{Closed superstring spectrum for the ground and the first excited levels of $NS$ sector and 
ground state of $R$ sector. ${\bf 1, 7, 27}$ are the tensor representations  
and ${\bf 8, 48}$ are the spinor representations of $SO(7)$. }
\end{table}

\end{document}